\documentstyle[preprint,prb,aps,epsf]{revtex}

\begin{document}

\title{\bf Two-dimensional oriented self-avoiding walks with parallel contacts}
\author{G.T. Barkema}
\address{Institute for Advanced Study, Olden lane, Princeton NJ 08540}
\author {S. Flesia
\thanks{present address: Dept. of Mathematics, Imperial College,
Huxley Building, London SW7 2BZ}}
\address{Theoretical Physics, University of Oxford,\\
1 Keble Road, Oxford, OX1 3NP, United Kingdom.}

\maketitle

\begin{abstract}
Oriented self-avoiding walks (OSAWs) on a square lattice are studied, with
binding energies between steps that are oriented parallel across a face of
the lattice. By means of exact enumeration and Monte Carlo simulation, we
reconstruct the shape of the partition function and show that this system
features a first-order phase transition from a free phase to a tight-spiral
phase at $\beta_c=\log(\mu)$, where $\mu =2.638$ is the growth constant for
SAWs. With Monte Carlo simulations we show that parallel contacts happen
predominantly between a step close to the end of the OSAW and another step
nearby; this appears to cause the expected number of parallel contacts to
saturate at large lengths of the OSAW.
\end{abstract}

\section{Introduction}
Many aspects of the behavior of polymers can be described by self-avoiding
walks on a lattice. Some polymers have interactions that depend on the
spatial orientation of the polymer, for instance A-B polyester.
Such polymers are conveniently modeled by {\it oriented} self-avoiding walks
(OSAW) with two types of short-ranged interaction between edges depending on
their relative orientation \cite{cardy94,debbieetal95,flesia95,koo95}.

The model of investigation in this paper consists of one OSAW on a
square lattice. Besides self-avoidance, the only interactions of the
OSAW with itself occur if two steps of the walk are one lattice spacing apart.
If the two steps have the same orientation, they are said to form a
{\it parallel contact}, to which an energy gain of $\epsilon_p$ is
attributed. If they have opposite orientation, they are said to form an
{\it anti-parallel contact}, with an energy gain of $\epsilon_a$.
If $\beta$ is the inverse temperature, and we define
$\beta_p=-\beta \epsilon_p$ and $\beta_a=-\beta \epsilon_a$, then
the partition function of such an oriented self-avoiding walk is given by
\begin{equation}
Z_n(\beta_p,\beta_a)=\sum_{m_p,m_a} C_n(m_p,m_a) e^{\beta_p m_p+\beta_a m_a}\;,
\end{equation}
where the sum is over all allowed values of the number of parallel
contacts, $m_p$, and the number of anti-parallel contacts, $m_a$, and
$C_n(m_p,m_a)$ is the number of configurations of length $n$ with $m_p$
parallel and $m_a$ anti-parallel contacts.
The limiting reduced free energy per step is given by
\begin{equation}
\label{freeenergy}
F(\beta_p,\beta_a)= \lim_{n\rightarrow\infty} \frac{1}{n}\log
\left[Z_n(\beta_p,\beta_a)\right]\;. 
\end{equation}

The phase diagram of this model has been studied previously 
\cite{debbieetal95}; numerical results from exact series up to $n=29$
edges showed the existence of three phases: a free SAW phase, a normal
collapsed phase and a compact spiral phase. The transition from
the free to the spiral phase was conjectured to be of first order.

In this article we will concentrate on the case where there are only 
interactions between parallel contacts, i.e. $\beta_a=0$.
The earlier work \cite{debbieetal95} rigorously proved that for this case
the reduced limiting free energy is constant for $\beta_p \leq 0$ with
value $\log (\mu)$, where $\mu$ is the growth constant for SAW $(\mu =2.638)$.
For $\beta_p>0$ the following rigorous bounds were proved:
\begin{equation}
\label{paralbound}
\beta_p \leq F(\beta_p,0) \leq \beta_p + \log (\mu).
\end{equation}
The above results prove the existence of a phase transition for
$0 \le \beta_p \leq \log (\mu)$. Bennett-Wood {\it et al} \cite{debbieetal95}
conjectured that
the critical inverse temperature $\beta_c$ is near or equal the lower bound
which, for $\beta_a =0$, is $\log (\mu) \approx 1$. In section \ref{phasetrans}
we further investigate this phase transition by extending the exact enumeration
data, by means of Monte Carlo results and combining them with some rigorous
results on tight spirals.

Another interesting question concerning OSAWs is the mean number of contacts.
One of us proved that the mean number of anti-parallel contacts
$\langle m_a \rangle \sim n$ in two or higher dimensions, where $n$
is the number of steps of the walk \cite{flesia95}. The mean number of
parallel contacts scales as $\langle m_p \rangle \sim n$ in
three or higher dimensions, but in two dimensions the behavior is still
an open question.
Field theoretic work \cite{cardy94} predicts that in two dimensions 
$\langle m_p \rangle \sim \log (n)$ in the limit $n \rightarrow \infty$.
However, Monte Carlo results for OSAWs with up to 3000 steps seem to indicate
that $\langle m_p \rangle$ tends to a constant $\approx 0.05$ \cite{flesia95}.
In section \ref{mp} we present the results of large-scale Monte Carlo
simulations with OSAWs of up to 5000 steps, and investigate these results
in a way that allows extrapolations to even larger $n$. Based on these
results we obtain an upper bound for $\langle m_p \rangle$ in the limit
$n \rightarrow \infty$.
 
\section{Phase transition towards a tight spiral}
\label{phasetrans}

Bennett-wood {\it et al} \cite{debbieetal95} enumerated all configurations
up to SAWs with a length of $n=29$ and ordered them according to their
number of parallel and anti-parallel contacts. We extended the exact
enumeration of the OSAWs with parallel contacts, and obtained all values
for $C_n(m_p)$, the number of OSAWs consisting of $n$ steps and having $m_p$
parallel contacts, up to $n=34$.

In our enumeration program, we start with generating all OSAWs of length
$l\leq n$ with a parallel contact between the first and the last step.
For each walk $w$, we determine the number of parallel contacts $m_w$.
We also determine $M_i(w,t_i,m_i)$, the number of extensions on the inside
end of walk $w$ with length $t_i \leq n-l$ and $m_i$ parallel contacts 
with either itself or $w$, and $M_o(w,t_o,m_o)$, the corresponding quantity
for the extensions on the outside end. Since the walk $w$ prevents contacts
between the inside and outside extensions, the total number of OSAWs of
length $n$ with $m_p$ parallel contacts is given by
\begin{equation}
Z_n(m_p)=\frac{1}{m_p} \sum_w \sum_{l+t_i+t_o=n}
          M_i(w,t_i,m_i)~~M_o(w,t_o,m_o)~~\delta(m_p,m_w+m_i+m_o).
\end{equation}
The prefactor in this equation corrects for the fact that there are $m_p$
different walks $w$ from which we can generate the same OSAW with
$m_p$ parallel contacts. Exploiting rotational and mirror symmetry, we 
enumerated all OSAWs of length $n \leq 34$ with one or more parallel contacts
in a run of about two weeks on a four-processor DEC alpha workstation.
Finally, the number of OSAWs without parallel contacts is obtained by
subtraction from the total number of OSAWs (from Ref. 7).

The results are given in table \ref{enumtab},
and plotted as the solid lines in figure \ref{exactenum}, where
$\log(C_n(m_p))$ is plotted as a function of $m_p$. The figure shows that up
to $n=34$, the number of configurations as a function of the number of
parallel contacts first drops quickly with a factor $p_n$, but then,
over the whole range $1 \leq m_p \leq m_{max}$, falls off exponentially
with the same exponent $q_n$. The partition function $Z_n(\beta_p)$ is thus
described well by 
\begin{eqnarray}
\label{partfunc}
\nonumber
C_n(1) & = & p_n C_n(0) \\
C_n(m) & \approx & C_n(1) \cdot \exp (-q_n (m-1)),
\end{eqnarray}
where $p_n$ and $q_n$ are $n-$dependent parameters.
 
To extend the graph presenting the partition function beyond $n=34$
by means of exact enumeration is very hard. However, the left
part of this graph for much larger $n$ can be obtained statistically by
means of Monte Carlo simulations: OSAWs are randomly generated with the
pivot algorithm \cite{pivot}, and a histogram is made of the number of
parallel contacts of these OSAWs. This gives us a direct measurement of
$C_n(m_p)/Z_n(0)$ for a small number of parallel contacts. In our Monte Carlo
simulations, we thermalized over $10^7$ pivot moves, followed by $10^8$
moves to gather statistics; statistical errors were obtained by repeating
the whole procedure 10 times. The results are shown in table \ref{pntable};
the density of OSAWs with more than $\sim 10$ parallel contacts is so small
that they will most likely never be generated, and we only obtain an upper
bound for them. A good approximation for $Z_n(0)$ is known:
\begin{equation}
\label{totnum}
Z_n(0) \approx (A/4) \mu^n n^{\gamma_s-1}
\end{equation}
where $\mu =2.638$, $\gamma_s=43/32$ \cite{gamma}, and $A=1.771$ \cite{Z_2_39}.
The factor of a fourth is due to the fact that we count OSAWs that are
equivalent after rotation only once. The Monte Carlo results from table
\ref{pntable} for $n=50, 60, 70, 80, 90$ and 100, multiplied by $Z_n(0)$,
are plotted as circles in the left side of figure \ref{exactenum}.

Also the utmost right point of the graph can be obtained, as there the only
relevant configurations are tight spirals. The corners of a tight spiral
are reached after $n=k,k+1,2k+2,2k+4,3k+6,3k+9, \cdots$ steps, {\it i.e.}, at
$n=ik+i^2$ or $n=ik+i(i-1)$, where $k$ is the number of steps in the same
direction at the inner end of the tight spiral, and $i$ is a positive integer.
Each additional step of the tight spiral adds one parallel contact, except
steps before and after a corner. Thus, the number of parallel contacts
$m_{max}$ for an OSAW of length $n$ is given by
\begin{eqnarray}
\nonumber
\mbox{if}~~(n \leq 2k):
& m_{max}=& 0; \\
\nonumber
\mbox{if}~~(n>2k):
& m_{max}=& n-2k+3-\left[\sqrt{n+\frac{ k   ^2}{4}  }-\frac{ k   }{2}\right]
                  -\left[\sqrt{n+\frac{ k   ^2}{4}-1}-\frac{ k   }{2}\right]\\
\label{tighteq}
&&                -\left[\sqrt{n+\frac{(k-1)^2}{4}  }-\frac{(k-1)}{2}\right]
                  -\left[\sqrt{n+\frac{(k-1)^2}{4}-1}-\frac{(k-1)}{2}\right],
\end{eqnarray}
where square brackets denote the Entier function ($[x]$ is the largest integer
not larger than $x$).
The number of parallel contacts of a `rectangular' tight spiral (with $k>1$)
does never exceed that of the `square' tight spiral (with $k=1$), but
can be equal, adding to the degeneracy of the ground state.
Additional ground states can be generated by removing steps from the inside
and adding them to the outside end, until the corner is reached.
Also, if the tight spiral ends at a corner or one or two steps further,
additional groundstates arise by rearranging these last steps.

We enumerated all OSAWs with $m_{max}$ parallel contacts and length up to
$n=50$, and confirmed that all groundstates can be generated with these
operations. Assuming that no new types of degenerate groundstates arise after
$n=50$, we calculated the degeneracy of the ground state for lengths
up to a million steps, and observed that the degeneracy fluctuates between 4
(for a complete `square' tight spiral) and $c_m n^{3/4}$ with $c_m=5.3$,
whereas the expected degeneracy grows as $c_a n^{3/4}$ with $c_a=2.1$.
For $n=50, 60, 70, 80, 90$ and 100, there are 140, 40, 16, 4, 16, and 8 
configurations with the maximum number of parallel contacts.  We have added
these results of the tight spirals in figure \ref{exactenum} as squares.

For $n\gg 34$, the Monte Carlo data in figure \ref{exactenum} for small $m_p$
do not extrapolate to the exact results for tight spirals, but point below,
which suggest that eq. (\ref{partfunc}) is an upper bound for $n\gg34$.
The dotted lines in figure \ref{exactenum} represent these upper bounds.
We cannot exclude the possibility that for $n\gg 34$ the partition function
initially stays below these dotted lines, then increases and crosses this
dotted line for intermediate values of $m_p$, and finally reaches the
exact result for tight spirals; however, we think that that scenario is
unlikely, and the results concerning long OSAWs in the remainder of
this section are based on the assumption that the dotted lines in figure
\ref{exactenum} represent upper bounds.

For $n \leq34$ we know $C_n(0)$ and $C_n(1)$ by exact enumeration, and
for $n=50,$ 60, 70, 80, 90, 100, 1000 and 2000 we know $C_n(0)/Z_n$ and
$C_n(1)/Z_n$ accurately from the Monte Carlo simulations. This enables us
to compute $p_n$ in eq. (\ref{partfunc}) for all these values of $n$. For
large $n$, $p_n$ converges to a constant value around 0.031. To extract
the specific heat and density of parallel contacts, we used a fit to $p_n$
which is given by
\begin{equation}
\label{pn}
p_n -p_{\infty}\sim (1/\sqrt{n}),
\end{equation}
where $p_{\infty}=0.031 \pm 0.002$. We can obtain the values $q_n$
in eq. \ref{partfunc} from equations (\ref{totnum}), (\ref{tighteq}), and
(\ref{pn}), as
\begin{equation}
\label{qn}
q_n \approx \frac{\log (C_n(1))-\log(C_n(m_{max}))}{m_{max}-1}
            \approx \frac{\log(Z_n)+\log(p_n)-3/4~\log(n)}{m_{max}-1}.
\end{equation}
For $n$=1000 and 2000, this equation predicts that $q_n$=1.099 and 1.060, 
respectively, whereas the Monte Carlo results in table \ref{pntable} for
$C_n(1)/C_n(5)$
indicate that the slope of $\log(C_n(m_p))$ corresponds to values of
$q_n \approx 1.4$; for larger values of $n$ the curves of $\log(C_n(m_p))$
versus $m_p$ initially point below the point corresponding to the
tight-spiral configuration, and thus must bend upwards at larger $m_p$.

For $n$ up to $34$ we plotted in figure \ref{specheat} the specific heat,
defined by $\beta \chi= -\partial^2 F/\partial \beta^2$, and in figure
\ref{mpdens} we plotted the density of parallel contacts
$\langle m_p \rangle/n$, as a function of the inverse temperature $\beta$. 
In both figures we added the graphs for $n=50$, 100, 200, 500, 1000, 2000
and 10,000, obtained from eq. (\ref{partfunc}), as dotted lines.
In figure \ref{specheat}, the value of $\beta$ where the peak of the
specific heat is located is moving backward to $\beta=\log(\mu)$, as
is the point where $\langle m_p \rangle /n $ is 
increasing steeply in figure \ref{mpdens}.
The jump in the density of parallel contacts (i.e., the energy density)
is increasing with increasing $n$, indicating a first order transition.
In fact, assuming eq. (\ref{partfunc}) one can show analytically that in the
limit $n \rightarrow \infty$ the function $\langle m_p \rangle /n$ approaches
the Heaviside stepfunction $\Theta(\log(\mu))$, and this still holds if eq.
(\ref{partfunc}) is an upper bound rather than an exact expression in the
regime between tight spirals and walks with few parallel contacts.
Both the specific heat and the density of parallel contacts are insensitive
to the fact mentioned earlier, that the curve starts somewhat steeper at
small $m_p$ and thus must bend up at larger $m_p$. If anything, they will
increase the peak value of the specific heat, and the steepness of the
density curve.

Another way to estimate the transition point is to look at the zeroes of the
partition function \cite{yang52,lee52}.
The partition function of an OSAW of $n$ steps with $m_p$ parallel contacts
is a polynomial of degree $m_{max}$ (the maximum number of parallel contacts)
in the variable $x=e^{\beta}$, hence it can be conveniently written in terms
of its $n$ roots ${r_{m_p}}$ in the complex plane:
\begin{equation}
Z_n(x)= C_n(0) \prod_{m_p=1}^{m_{max}} (1-(x/r_{m_p}))
\end{equation}
and the free energy per steps
\begin{equation}
F_n(x)=\frac{1}{n} \log (C_n(0)) + \frac{1}{n} \sum_{m_p=1}^{m_{max}} 
\log (1-(x/r_{m_p})).
\end{equation}
The coefficients $C_n(m)$ are real and non-negative, hence none of the 
roots lies on the real positive axis, but for $n \rightarrow \infty$
they will cross it at some point $x_c \le \mu$, since we rigorously know the 
existence of a phase transition.

We calculated the zeroes of the partition function corresponding to the 
exact data up to $n=34$ and they are plotted in figure \ref{zeroes}a.
The roots seem to lie in nearly perfect circles for every $n$, but the 
radius decreases with increasing $n$. The $n^{th}$ roots nearest to the real
positive axis approach the real axis along a nearly straight line.
In figure \ref{zeroes}b, we plotted the real part of the root nearest to
the real axis for $n=25..34$, against $1/n$. 
Again, the figures are consistent with a transition at $x_c \approx 2.5$.

\section{Number of parallel contacts for $\beta=0$}
\label{mp}

The second major topic of this paper is to investigate the behavior of the
number of parallel contacts $m_p$ in the limit $n \rightarrow \infty$.
In figure \ref{mpplot} we have plotted the behavior of $\langle m_p \rangle$
as a function of $n$, obtained from eq. (\ref{partfunc}), which we proposed
to be an upper bound.  The upper bound reaches asymptotically the value
$m_p=0.08$. Clearly, the earlier mentioned fact that the
curve has a somewhat steeper slope at large $n$ and small $m_p$ has impact
on $\langle m_p \rangle$, as these configurations are dominant at $\beta=0$.
Therefore we do not use eq. (\ref{partfunc}) in the remainder of this section.
With Monte Carlo simulations we have determined the expected number
of parallel contacts $\langle m_p \rangle$ as a function of $n$. The
results are given in table \ref{mpexp} and figure \ref{mpplot}, and are
in agreement with results published earlier by one of us \cite{flesia95},
but extend to larger values of $n$.
The Monte Carlo results seem to converge to a value around 0.05.

To understand the underlying physics in the regime $\beta=0$ better,
we took a closer look on {\it where} the parallel contacts are made, and 
relate this to other types of SAWs. Consider
an oriented OSAW of length $n$, with a parallel contact between the steps
$i$ and $j$ of the walk. The sequence of steps from $i$ to $j$ constitute a
polygon of length $l=j-i+1$, if one of the two steps that form a contact is
rotated 90 degrees to close the polygon. The remaining sequences of steps
from $0$ to $i$ and from $j$ to $n$ are two self-avoiding walks of length
$i$ and $n-j$, respectively. These two SAWs can be combined into one
self-avoiding two-legged star: a SAW of length $n-l$, on which one special
point (the origin of the two-legged star) is marked. Note that, since the two
SAWs are separated by the loop, one being located on the inside of the loop
and one on the outside, the two-legged star is always self-avoiding.
The mapping of an OSAW with one parallel contact into a rooted polygon
plus a two-legged star is illustrated in figure \ref{twolegged}.

If an OSAW has more than one parallel contacts, then we can map this OSAW
onto different combinations of a rooted polygon plus a two-legged star.
In general, if the OSAW has $m_p$ parallel contacts, there are $m_p$ such
mappings into a rooted polygon plus a two-legged star.  The reverse mapping,
i.e. the combination of a two-legged star plus a rooted polygon into an OSAW
with a parallel contact, is not guaranteed to result in an OSAW with a
parallel contact, as they might cross. Therefore, the total number of
rooted polygons of length $l$ times the total number of two-legged stars
of length $n-l$, summed over all $l$, is an upper bound to the number of
OSAWs of length $n$, multiplied by the expectation value of the number of
parallel contacts for these walks.

Let us define $f(n,l)$ as the probability that a two-legged star of length
$n-l$ if combined with a rooted polygon of length $l$, results in an OSAW.
Then we can write
\begin{equation}
\label{expm}
\langle m_p \rangle Z_n=\sum_{m_p} m_p C_n(m_p)=\sum_l P_l S_{n-l} f(n,l)
\end{equation}
where $Z_n$, $P_n$ and $S_n$ are the number of OSAWs, rooted polygons and
two-legged stars of length $n$, respectively.

We know that, for large $n$:
\begin{eqnarray}
Z_n\approx \mu^n n^{\gamma_s-1}  \\
S_n\approx \mu^n n^{\gamma_s}  \\
P_n\approx \mu^n n^{\alpha-2}
\end{eqnarray}

Combining this with (\ref{expm}) leads to:

\begin{equation}
\langle m_p \rangle = \sum_l
\frac {l^{\alpha-2} (n-l)^{\gamma_s} f(n,l)} {n^{\gamma_s-1}}
\end{equation}

We can obtain insight in the behavior of the function $f(n,l)$ by means of
Monte Carlo simulations. OSAWs are sampled randomly, and for each parallel
contact the loop length $l=|j-i+1|$ is determined, where $i$ and $j$ are the
steps making the parallel contact. This procedure gives us 
$\langle m_p \rangle (l)$, the expectation value of the number of parallel
contacts with loop length $l$.  Results for OSAWs with a length of
$n=$200, 500, 1000, 2000, and 5000 are plotted in figure \ref{mploop}.
$\langle m_p \rangle (l)$ shows a power-law behavior, where
the length $n$ of the OSAW is an upper bound to the length $l$ of the loop.
Important however is that, besides this obvious dependence, the total
length $n$ does not appear to have any influence on the behavior of
$\langle m_p \rangle (l)$, and this quantity is well described by a power-law:
\begin{eqnarray}
\langle m_p \rangle (l) \approx k~~l^{-\alpha_l}
\end{eqnarray}
Numerically, we find:
\begin{eqnarray}
k & = & 0.35 \pm 0.1 \\
\alpha_l & = & 1.65 \pm 0.05
\end{eqnarray}
To obtain the mean number of parallel contacts $\langle m_p \rangle$ we sum 
over all possible (even) lengths $l$ of the rooted polygon:
\begin{equation}
\label{mln}
\langle m_p \rangle = \sum_l \langle m_p \rangle (l) \approx
k~~ \sum_{l=8}^n l^{-\alpha_l}.
\end{equation}
For $n\rightarrow \infty $ the right hand side equals a constant times the
function $\zeta(\alpha_l)$, which converges to a constant for $\alpha_l>1$.
This implies again that $\langle m_p \rangle$ tends to a constant in agreement
with earlier Monte Carlo results of Flesia \cite{flesia95}.

The fact that $\langle m_p \rangle$ is constant implies that the SAW
critical exponent $\gamma$ is constant in the free and repulsive regime
(i.e. for $\beta \le 0$), and presumably until the transition.
For the exponent $\gamma$ to change with $\beta$,
the exponent $\alpha_l$ should be $\le 1$, since this will cause
the $\zeta$ function to diverge, but this is not supported by our
numerical results in Fig. 7.

It is possible to put an upper limit to how far $\langle m_p \rangle$ will
still increase if $n$ is increased above 5000: figure \ref{mploop} shows
that the contribution of loops with a length below 1000 certainly has
converged for $n=5000$, thus $\langle m_p \rangle (\infty)-
\langle m_p \rangle (5000) < k \cdot \sum_{l=1000}^n l^{-\alpha_l} < 10^{-5}$.

A different approach which estimates both the number of parallel and
anti-parallel contacts is to use the similarity between an OSAW and a
twin-tailed tadpole.  Consider an OSAW with a contact between steps $i$
and $j$ of the walk.  If we add a new edge between steps $i$ and $j$ we
obtain an object which we will call a {\it twin-tailed loop} (see figure 8).
A twin-tailed loop differs from a non-uniform twin-tailed tadpole only by
one edge, and has the same asymptotic behavior.  If the contact is parallel
then the twin tailed loop has one tail inside the loop and the other outside
(see fig 8a), while if the contact is anti-parallel both tails are outside
(see fig 8b). This is of course only true in two dimensions. Each OSAW with
$m$ contacts can be mapped into $m$ distinct twin-tailed loops. If $T_n$ is
the total number of twin-tailed loops of total length $n$ then it follows that
\begin{equation}
T_n =\sum \left( m~C_n(m) \right).
\end{equation}
Dividing both sides by $Z_n$, where $Z_n$ is the partition function of SAWs,
it follows that
\begin{equation}
\label{ttt}
\langle m \rangle=T_n/Z_n
\end{equation}
Asymptotically, $Z_n \sim \mu^n n^{\gamma_s -1}$, where $\gamma_s$ is the
exponent for SAWs. Lookmann~\cite{lookmann} proved that twin-tailed tadpoles
have the same growth constant $\mu$ as SAWs and that the exponent $\gamma$
is $\gamma=\gamma_s+1$. The same kind of proof holds for twin-tailed loops.
Replacing these results in eq. (\ref{ttt}) implies the known result 
$\langle m \rangle \sim n$.

Consider now the parallel and the anti-parallel case separately.
Twin-tailed loops with both tails outside the loop are the dominant
configurations, so they have the exponent $\gamma$ of the total set,
{\it i.e.} $\gamma=\gamma_s+1$. This implies as previously that
$\langle m_a \rangle \sim n$ as was proven by one of us~\cite{flesia95}.

Parallel contacts correspond to the subset $T^*_n$ of twin-tailed loops with
one tail on the inside and one on the outside of the loop. The question is,
what is the value of the exponent $\gamma$ (let us call this exponent
$\gamma_t$) for this subset $T^*_n$? Simple tadpoles ({\it i.e.} tadpoles
with only one tail) have the same $\gamma$ as SAWs~\cite{lookmann}.
Since one element of $T^*_n$ can be constructed from a simple tadpole
by adding one edge inside the loop, it follows that $\gamma_t \geq \gamma_s$.
On the other hand, since $T^*_n$ is a subset of the set of twin-tailed loops,
it follows that $\gamma_t \leq \gamma_s +1$, and this inequality can be made
strict by considering that $\langle m_p \rangle \sim o(n)$, see
Bennett-Wood {\it et al.}\cite{debbieetal95}.

We can gain insight in this matter by randomly generating OSAWs of length $n$,
and for each parallel contact determining the length $t$ of the inside tail.
Note that if a parallel contact is formed between steps $i$ and $j$ of the
OSAW, the steps from $i$ to $j$ form a loop, and `inside' and `outside' tails
refer to inside or outside this loop. The results are plotted in figure 9.
Extrapolating these results we estimate that the fraction of twin-tailed
loops with length $t$ of the inside tail is decreasing as  
\begin{equation}
\label{inside}
\langle m_p \rangle (t) \sim k_t t^{-\alpha_t}
\end{equation}
where $\alpha_t=1.6 \pm 0.1$. The parameters $\alpha_l$ and $\alpha_t$
are within each others statistical errors and are probably the same.
As in Eq. (\ref{inside}) the parameter $\alpha_t$ exceeds 1, $\sum_t (m_p(t))$
will not be more than a constant times $m_p(t=0)$. This implies that 
$T^*_n$ asymptotically seems to behave as simple tadpoles
which have the same $\gamma$ as SAWs.  If we assume, based on these numerical
results and intuitive arguments, that the twin-tailed loops with one tail
inside and one outside behave as simple tadpoles then $\gamma_t =\gamma_s$,
which would imply that $\langle m_p \rangle$ approaches a constant. 
 
\section*{Acknowledgements}

We like to thank Alan Sokal, John Cardy, John Wheater, and Stu Whittington
for fruitful discussions.
G.T.B. acknowledges financial support from the EPSRC under Grant No. GR/J78044,
from the DOE under Grant No. DE-FG02-90ER40542, and from the Monell Foundation.
S.F. is grateful to EPSRC of U.K. for financial support (grant B/93/RF/1833).

\begin{table}[htb]
\caption{exact enumeration of the number of OSAWs of $n$ steps, with $m_p$
parallel contacts.
\label{enumtab}}
\begin{tabular}{llllllllllllllll}
n & $m_p$=0 & 1 & 2 & 3 & 4 \\
\hline
30 &
4173469695963 &
61649050972 &
8921988104 &
1417268612 &
221155744 \\

31 &
10975225680123 &
163203273852 &
25422408744 &
3820038428 &
663920466 \\

32 &
29224474453695 &
453395153136 &
67676366244 &
11044497696 &
1800473376 \\

33 &
77923458322683 &
1201209580824 &
190907785004 &
29775283928 &
5291859172 \\

34 &
207390873801535 &
3318007864896 &
508582438722 &
84979159776 &
14355126160 \\
\hline\hline

n & 5 & 6 & 7 & 8 & 9\\
\hline
30 &
35795108 &
5383888 &
801432 &
108062 &
16652 \\

31 &
98665196 &
17463042 &
2253640 &
399888 &
46368 \\

32 &
301423940 &
48238616 &
7546064 &
1123840 &
177756 \\

33 &
830969056 &
150009218 &
21332880 &
3819684 &
510908 \\

34 &
2474324280 &
415293124 & 
67773784 &
10824900 &
1773072 \\
\hline\hline

n & 10 & 11 & 12 & 13 & 14 & 15\\
\hline
30 &
1372 &
272 &
16 \\

31 &
7188 &
640 &
164 \\

32 &
20000 &
3512 &
332 &
48 \\

33 &
81240 &
10096 &
1976 &
72 &
28 \\

34 &
235146 &
40728 &
5294 &
704 &
40 &
16 
\end{tabular}

\end{table}

\begin{table}[htb]
\caption{Monte Carlo results for the density of OSAWs of length $n$ with
$m_p$ parallel contacts.
\label{pntable}}
\begin{tabular}{llllllllllllllll}
n & $m_p$=0 & 1 & 2 & 3 & 4 \\
\hline
50 &
0.97763(1) &
0.01841(1) &
0.003209(8) &
0.000599(3) &
0.000120(1) \\
60 &
0.97603(2) &
0.01954(2) &
0.003555(7) &
0.000696(3) &
0.0001426(8) \\
70 &
0.97479(3) &
0.02039(3) &
0.003832(7) &
0.000780(4) &
0.000164(1) \\
80 &
0.97368(2) &
0.02118(2) &
0.004067(10) &
0.000840(6) &
0.000180(2) \\
90 &
0.97280(3) &
0.02178(2) &
0.00426(1) &
0.000903(3) &
0.000198(2) \\
100 &
0.97210(2) &
0.02229(2) &
0.00441(1) &
0.000933(4) &
0.000209(2) \\
1000 &
0.9629(4) &
0.0284(3) &
0.0066(1) &
0.00159(7) &
0.00042(2) \\
2000 &
0.9618(4) &
0.0293(5) &
0.0067(1) &
0.00166(5) &
0.00043(4) \\
\hline\hline
n & 5 & 6 & 7 & 8 & 9 \\
\hline
50 &
0.0000233(4) &
0.0000052(3) &
0.00000112(9) &
0.00000012(3) &
0.00000002(2) \\
60 &
0.0000300(5) &
0.0000062(2) &
0.0000013(1) &
0.00000026(5) &
0.00000008(2) \\
70 &
0.0000346(6) &
0.0000080(3) &
0.0000015(2) &
0.00000021(4) &
0.00000009(4) \\
80 &
0.0000401(8) &
0.0000087(5) &
0.0000019(2) &
0.00000032(7) &
0.00000014(3) \\
90 &
0.000044(1) &
0.0000109(3) &
0.0000020(2) &
0.00000045(6) &
0.00000011(4) \\
100 &
0.000047(1) &
0.0000112(5) &
0.0000029(2) &
0.0000007(1) &
0.00000012(4) \\
1000 &
0.000102(8) &
0.000032(6) &
0.000009(2) &
0.0000007(4) &
\\
2000 &
0.00012(2) &
0.000024(5) &
0.000009(3) &
0.0000010(4) &
&
\end{tabular}
\end{table}

\begin{table}[htb]
\caption{Monte Carlo data for $\langle m_p \rangle$, the expected number
of total parallel contacts.
\label{mpexp}}
\begin{tabular}{llll}
n & $\langle m_p \rangle$ & n & $\langle m_p \rangle$ \\
\hline
9 & 0.001966(3) & 10 & 0.00505(3) \\
11 & 0.00450(5) & 12 & 0.00715(3) \\
13 & 0.00698(2) & 14 & 0.00918(3) \\
15 & 0.00921(3) & 16 & 0.01106(4) \\
17 & 0.01118(3) & 18 & 0.01274(4) \\
19 & 0.01293(2) & 20 & 0.01429(3) \\
21 & 0.01446(4) & 22 & 0.01577(4) \\
23 & 0.01592(2) & 24 & 0.01693(7) \\
28 & 0.01925(2) & 29 & 0.01953(7) \\
30 & 0.02025(3) & 38 & 0.02358(5) \\
39 & 0.02389(7) & 40 & 0.02431(8) \\
41 & 0.02448(4) & 48 & 0.02667(5) \\
49 & 0.02690(8) & 50 & 0.02731(7) \\
70 & 0.03129(7) & 71 & 0.03141(5) \\
80 & 0.03281(7) & 90 & 0.0338(4) \\
99 & 0.0350(5) & 120 & 0.0372(4) \\
150 & 0.0385(4) & 200 & 0.0406(4) \\
300 & 0.0429(4) & 400 & 0.0446(4) \\
500 & 0.0462(7) & 700 & 0.0471(6) \\
1000 & 0.0492(9) & 1500 & 0.0493(8) \\
2000 & 0.0497(8) & 3000 & 0.0497(9) \\
5000 & 0.0514(3) &
\end{tabular}
\end{table}

\begin{figure}
\caption{
A graphical representation of the partition function: the logarithm of
the number of OSAWs is plotted as a function of its number of parallel
contacts. Solid lines are data for $n=11..34$, obtained from exact
enumeration, circles are data for $n=50$, 60, 70, 80, 90, and 100, obtained
from Monte Carlo simulations, squares from properties of tight spirals,
and the dotted lines are connecting the Monte Carlo results with the
corresponding results for tight spirals.
\label{exactenum}}
\end{figure}

\begin{figure}
\caption{Specific heat as a function of inverse temperature $\beta$.
In the direction of increasing peak value, the curves are obtained
for $n$=25, 30, and 34 from exact enumeration (solid lines) and for $n$=50,
100, 200, 500, 1000, 2000 and 10000 from eq. (4) (dashed lines).
\label{specheat}}
\end{figure}

\begin{figure}
\caption{density of parallel contacts, as a function of inverse temperature
$\beta$. In the direction of increasing density, the curves are obtained
for $n$=25, 30, and 34 from exact enumeration (solid lines) and for $n$=50,
100, 200, 500, 1000, 2000 and 10000 from eq. (4) (dashed lines).
\label{mpdens}}
\end{figure}

\begin{figure}
\caption{Left: zeroes of the polynomial of the partition function for
$n=25..34$. Right: The zeroes of the $n$-th root approach the real axis
nearly along a straight line, crossing the real axis at $x_c \approx 2.5$.
\label{zeroes}}
\end{figure}

\begin{figure}
\caption{Expected number of parallel contacts, as a function of length.
The circles with error bars are Monte Carlo measurements, the solid line 
results from eq. (4) and is an upper bound.
\label{mpplot}}
\end{figure}

\begin{figure}
\caption{Decomposition of an OSAW into a loop and a two-legged star
\label{twolegged}}
\end{figure}

\begin{figure}
\caption{probability that an OSAW has a parallel contact with a loop of
length $l$, for OSAWs with a total length $n$=500, 1000, 2000 and 5000.
For each parallel contact, the loop length $l$ is defined as $l=|j-i+1|$,
where $i$ and $j$ are the steps of the OSAW making a parallel contact.
\label{mploop}}
\end{figure}

\begin{figure}
\caption{An OSAW with a contact can be transformed into a twin-tailed loop
by adding one step. If the contact was parallel, the twin-tailed loop has
one tail on the inside and one on the outside of the loop (see figure $a$).
If the contact was anti-parallel, both tails are located on the outside of the
loop (see figure $b$).
\label{twintail}}
\end{figure}

\begin{figure}
\caption{probability that an OSAW has a parallel contact with an inside tail
of length $t$, for OSAWs with a total length $n$=500, 1000, 2000 and 5000.
For each parallel contact, the steps $i$ up to $j$ form a loop, where $i$ and
$j$ are the steps of the OSAW making a parallel contact. The inside tail is
defined as those steps of the OSAW that are located within this loop.
\label{mptail}}
\end{figure}

\end{document}